\documentclass[twocolumn]{aastex631}

\usepackage{verbatim}
\usepackage{textcomp, gensymb}
\usepackage[normalem]{ulem}

\shorttitle{Proton and Electron Heating Rates}
\shortauthors{Bandyopadhyay et al.}
\graphicspath{{./}{figures/}}

\begin{document}

\title{Estimates of Proton and Electron Heating Rates Extended to the Near-Sun Environment}



\author[0000-0002-6962-0959]{R. Bandyopadhyay}
\affiliation{Department of Astrophysical Sciences, Princeton University, Princeton, NJ 08544, USA}

\author[0009-0006-2781-3484]{C. M. Meyer}
\affiliation{Department of Astrophysical Sciences, Princeton University, Princeton, NJ 08544, USA}

\author[0000-0001-7224-6024]{W. H. Matthaeus}
\affiliation{Department of Physics and Astronomy, University of Delaware, Newark, DE 19716, USA}

\author[0000-0001-6160-1158]{D. J. McComas}
\affiliation{Department of Astrophysical Sciences, Princeton University, Princeton, NJ 08544, USA}

\author[0000-0002-3699-3134]{S. R. Cranmer}
\affiliation{Laboratory for Atmospheric and Space Physics, Department of Astrophysical and Planetary Sciences, University of Colorado, Boulder, CO 80303, USA}

\author[0000-0001-5258-6128]{J. S. Halekas}
\affiliation{Department of Physics and Astronomy, University of Iowa,Iowa City,IA 52242,USA}

\author[0000-0002-9954-4707]{J. Huang}
\affiliation{Space Sciences Laboratory, University of California, Berkeley, CA 94720-7450, USA}

\author[0000-0001-5030-6030]{D. E. Larson}
\affiliation{Space Sciences Laboratory, University of California, Berkeley, CA 94720-7450, USA}

\author[0000-0002-0396-0547]{R. Livi}
\affiliation{Space Sciences Laboratory, University of California, Berkeley, CA 94720-7450, USA}

\author[0000-0003-0519-6498]{A. Rahmati}
\affiliation{Space Sciences Laboratory, University of California, Berkeley, CA 94720-7450, USA}

\author[0000-0002-7287-5098]{P. L. Whittlesey}
\affiliation{Space Sciences Laboratory, University of California, Berkeley, CA 94720-7450, USA}

\author[0000-0002-7728-0085]{M. L. Stevens}
\affiliation{Smithsonian Astrophysical Observatory, Cambridge, MA 02138, USA}

\author[0000-0002-7077-930X]{J. C. Kasper}
\affiliation{BWX Technologies, Inc., Washington DC 20002, USA}
\affiliation{Climate and Space Sciences and Engineering, University of Michigan, Ann Arbor, MI 48109, USA}

\author[0000-0002-1989-3596]{S. D. Bale}
\affil{Physics Department, University of California, Berkeley, CA 94720-7300, USA}
\affil{Space Sciences Laboratory, University of California, Berkeley, CA 94720-7450, USA}

 
\begin{abstract}
A central problem of space plasma physics is how protons and electrons are heated in a turbulent, magnetized plasma. The differential heating of charged species due to dissipation of turbulent fluctuations plays a key role in solar wind evolution. Measurements from previous heliophysics missions have provided estimates of proton and electron heating rates beyond 0.27 au. Using Parker Solar Probe (PSP) data accumulated during the first ten encounters, we extend the evaluation of the individual rates of heat deposition for protons and electrons in to a distance of 0.063 au $(13.5\,R_{\odot})$, in the newly formed solar wind. The PSP data in the near-Sun environment show different behavior of the electron heat conduction flux from what was predicted from previous fits to Helios and Ulysses data. Consequently, the empirically derived proton and electron heating rates exhibit significantly different behavior than previous reports, with the proton heating becoming increasingly dominant over electron heating at decreasing heliocentric distances. We find that the protons receive about $80 \%$ of the total plasma heating at $\approx 13\,R_{\odot}$, slightly higher than the near-Earth values. This empirically derived heating partition between protons and electrons will help to constrain theoretical models of solar wind heating.
\end{abstract}


\section{Introduction} \label{sec:intro}
The solar wind is a supersonic, magnetized plasma that flows into interplanetary space from the Sun. Observational evidence suggests continuous heat deposition into the solar wind plasma that begins in the corona and extends into the interplanetary space~\citep{Coleman1968ApJ, Leer1982SSR_acceleration, Tu1988JGR_heating, Grall1996Nature_acceleration, Verscharen2019LRSP}.
A leading candidate of this gradual heating is the dissipation of turbulent fluctuating energy that exists at large scales in the solar wind~\citep{Breech2009JGR_heating, VerdiniEA10}. To better understand coronal and solar wind heating, solar wind acceleration, and the large-scale evolution of the solar wind plasma, we need to know how energy is dissipated from turbulent fluctuations into different charged species.

The solar wind is a weakly collisional plasma, with the constituent charged species (e.g., protons, electrons, and heavy ions) often exhibiting significantly different temperatures, anisotropies, and even velocities~\citep{Marsch2006LRSP}. These deviations from thermal equilibrium are strongest in regions of low density and high temperatures where Coulomb collisions are infrequent~\citep[e.g.,][]{Neugebauer1982SSR_measurement, KasperEA08}. Relative to the other regions of the heliosphere, the fast solar wind exhibits very low density and very high temperature, and consequently very infrequent Coulomb collisions. These properties make the fast wind an optimal `plasma laboratory' for studies of collisionless kinetic processes associated with turbulent dissipation.

Many previous works have used the measurement of plasma properties in the solar wind to derive the rates of energy dissipation from processes such as magnetohydrodynamic (MHD) turbulence~\citep[e.g.,][]{Tu1988JGR_heating, VermaEA95, Smith2001JGR_heating, MacBride2005SW, Livadiotis2020ApJ_heating, Marino2022PR_scaling}. Most of these studies, however, analyzed the energy budget of a single charged species and did not consider protons and electrons together~\citep{Hellinger2011JGR_revisited, Stverak2015JGR_electron, Scudder2015ApJ_proton}. 

In an important antecedent to the present work, 
\citet{Cranmer2009ApJ_empirical} computed the individual estimates of proton and electron heating rates in the fast solar wind. This work used Helios and Ulysses data to estimate the heating rates from 0.3 au to about 5 au. A related investigation by~\citet{Breech2009JGR_heating} showed that with the value of heating partition given by~\citet{Cranmer2009ApJ_empirical}, a MHD turbulent heating model can also be consistent with the empirical data. The present paper makes substantial extension of this approach by employing newly available data much closer to the sun. 

NASA's Parker Solar Probe (PSP)~\citep{Fox2016SSR, Raouafi2023SSR_PSP} provides the first opportunity to extend these studies to the near-Sun environment, where the solar wind is `young' in its evolution and the physical conditions are very different~\citep{DeForestEA16, Chhiber2019ApJS_critical, Bandyopadhyay2021ApJ_geometry}. Some PSP studies have estimated the turbulence heating rate near the Sun, but most of them have not treated the energetics of the protons and electrons together~\citep{Bandyopadhyay2020ApJS_cascade, Sasikumar_Raja2021ApJ_heating, Abraham2022ApJ_Electrons, Wu2023PoP_review}. Although the mass density and momentum of the solar wind plasma are dominated by protons, the electron internal energy budget constitutes approximately half of the total plasma internal energy and can be rather important in its dynamics. A quantitative estimate of the heating partition between the two charged species is useful not only for accurate solar wind modeling, but also for constraining energy dissipation mechanisms in the young solar wind and the solar corona~\citep{Adhikari2021AA_model}. Therefore, a full treatment of solar wind plasma energetics should include both protons and electrons in the analysis but has not yet been reported close to the sun, as far as we are aware~\citep[however, see][]{Shankarappa2023arXiv_landau}.  Here, we use PSP data accumulated from the first 10 encounters alongside previous Helios and Ulysses data to estimate the individual rates of proton and electron heating from  approximately 0.06 au ($13\,R_{\odot}$), out to 5 au.

\section{In-Situ Data} \label{sec:data}
We focus on proton and electron plasma properties for the high-speed solar wind. PSP measurements within the solar encounter phase are used for distances closer than 0.25 au. We use data from the first ten solar encounters of PSP. The SWEAP~\citep{Kasper2016SSR} instrument suite provides measurements of proton temperature $T_p$ and outflow speed $u$ data. We utilized the moments data from Solar Probe Cup (SPC)~\citep{Case2020ApJS_SPC} and Solar Probe ANalyzers-Ion~\citep{Livi2022ApJ_SPANI} (SPAN-I) to obtain proton temperature and velocity, whenever the distribution function was within the field-of-view (FOV) of either instrument~\citep{Kasper2019Nature_SWEAP, Kasper2021PRL_corona}. No Maxwellian fits were used to derive the moments in this study. We have used SPC data for the first three encounters and SPAN-i data for the remaining seven encounters~\citep{Perez2021AA_Taylor}. The SWEAP measurements are further cleaned based on several criteria: we discard SPC data when the ``general flag" variable is
on; we check whether the peak of proton velocity distribution function (VDF) is within the FOV of SPAN-I, such as the solar wind flow angle is under an azimuthal angle of $\sim 165$ degrees in the instrument coordinate and the measured energy flux is peaked at least the second to the last azimuthal angle~\citep[e.g.,][]{Huang2023ApJ_switchback}.
Electron heat conduction flux $q_{||,e}$ data were obtained from the Solar Probe ANalyzers-Electrons (SPAN-E)~\citep{Whittlesey2020ApJS_SPANE} instrument of the SWEAP experiment, as described in~\citet{Halekas2020ApJS_electrons, Halekas2021AA_electrons}. FIELDS~\citep{Bale2016SSR} instrument suite measures magnetic and electric fields. Electron temperature $T_e$ and electron density $n_e$ data were evaluated from the quasi-thermal noise (QTN) spectrum provided by the electric field antennas~\citep{Pulupa2017JGR_space,Moncuquet2020ApJS_qtn,Martinovic2022JGR_plasma}.

\textit{Helios} and \textit{Ulysses} temperature and electron heat conduction data were taken from~\citet{Cranmer2009ApJ_empirical}. We utilize the assumption of 5\% helium abundance with $n_e = 1.1\;n_p$ to calculate $n_p$ and $n_e$ indirectly for the \textit{Helios} and \textit{Ulysses} proton density data and PSP electron density data, respectively.

We select the fast solar wind to be streams faster than 600 km s$^{-1}$ for all three datasets of PSP, Helios, and Ulysses. This high cutoff helps to minimize contamination from the slow solar wind, which often exhibits different properties than the fast wind~\citep{Cranmer2009ApJ_empirical,DassoEA05,MacBrideEA08}. 

\begin{figure}[t!]
    \centering
    \includegraphics[width=0.4\textwidth]{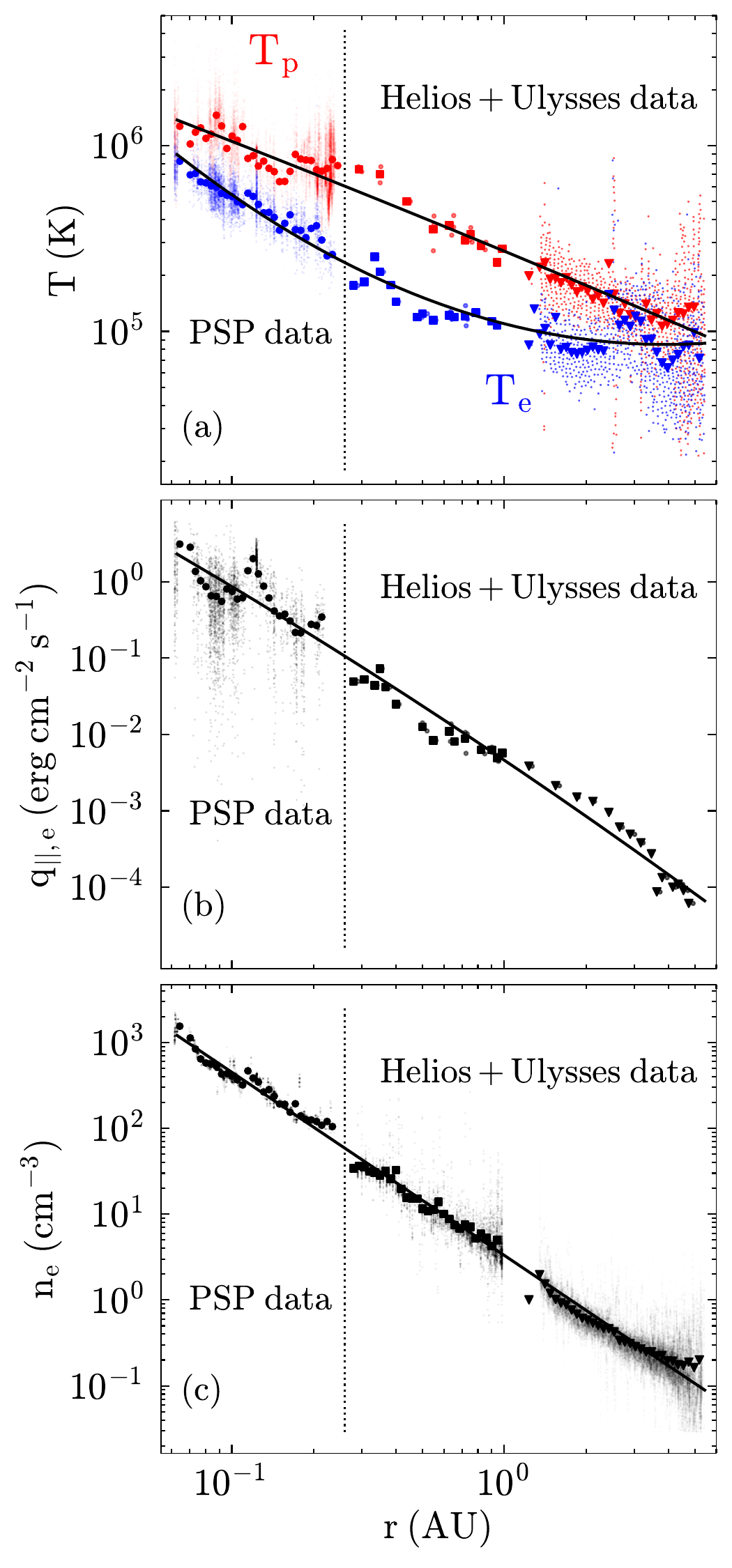} 
    \caption{\textit{In situ} measurements for the fast solar wind: (a) plasma temperature, (b) electron heat conduction flux, and (c) electron density. Binned \textit{PSP} (dots), \textit{Helios} (squares), and \textit{Ulysses} (triangles) data are shown in all plots, with protons in red and electrons in blue. Whenever binning has been performed, we have plotted the original data points in smaller, transluscent symbols. } \label{fig:1}
\end{figure}

\subsection{Analytic Fits} \label{subsec:fits}
Figure~\ref{fig:1} shows the radial dependence of the 
data along with the following analytic fits from the combined \textit{PSP}, \textit{Helios}, and \textit{Ulysses} datasets:

\begin{equation} \label{eq:1}
\text{ln} \left(\frac{T_p}{10^5\;\text{K}}\right) = 0.9939 - 0.6073 x - 0.0070 x^2,
\end{equation}

\begin{equation} \label{eq:2}
\text{ln} \left(\frac{T_e}{10^5\;\text{K}}\right) = 0.0941 - 0.3765 x + 0.1381 x^2,
\end{equation}
\begin{equation} \label{eq:3}
\text{ln} \left(\frac{q_{||,e}}{q_0}\right) = - 0.7752 - 2.4069 x - 0.0574 x^2,
\end{equation}
\begin{equation} \label{eq:4}
n_e = 3.3120\;r^{-2.1340},
\end{equation}
where $x =$ ln(r/[1 au]) and $q_0 = $0.01 erg cm$^{-2}$ s$^{-1}$. These fits are roughly consistent with the heat flux constraint in the collisionless limit~\citep{Bale2013ApJL_heat-flux}. We bin each individual dataset in the fitting process, thereby largely erasing the effect of non-uniformities in the number of data points. We notice that although the proton and electron temperature fits are rather close to the previous study by~\citet{Cranmer2009ApJ_empirical}, the radial fit for the electron parallel heat conduction flux differs in a significant way from \cite{Cranmer2009ApJ_empirical}. With the Helios and Ulysses data, the electron heat conduction  included data only for $r>0.29$ au, 
resulting in a more concave shaped fit near the inner 
boundary of that region. This causes the local slopes 
of the $q_{||,e}$ fits to be shallower approaching Helios perihelion, thus biasing the  fit towards a shallower slope in the inner heliosphere. The PSP electron heat flux data, however, continues to steeply increase at smaller radial distances, resulting in the very different fit seen in Fig.~\ref{fig:1}(b), 
compared to~\citet{Cranmer2009ApJ_empirical}.
As we see presently, these new fits cause significantly different behavior in the derived proton and electron heating rates.

\section{Results: Empirical Heating Rates} \label{sec:results}
Next, we use the obtained fits in the internal energy equations to estimate the proton and electron heating rates. In employing the internal energy conservation equations,
we neglect temperature anisotropies, which have been shown to affect the net heating only negligibly~\citep{Pilipp90JGR_variations,MatteiniEA07,VasquezEA07-cascade} and tend to return to isotropy as a result of plasma instabilities~\citep{Kasper2002GRL,HellingerEA06}. Further, there is some indication that proton temperature anisotropy may become weak close to the Sun in the regions explored by PSP~\citep{Huang2020ApJS_PSP, Cranmer2020RNAAS_updated}. We have also neglected proton heat conduction which is generally considered small
\citep{Braginskii65,Sandbaek95POP_transport},
while the contribution from electron heat conduction is larger and retained.

Therefore, we may write the 
steady state proton and electron internal energy conservation equations~\citep{Arya1991JGR_helios, Cranmer2009ApJ_empirical} in the following form:
\begin{eqnarray} \label{eq:5}
Q_p &=& \frac{3}{2} n_p u k_B \frac{\partial T_p}{\partial r} - u k_B T_p \frac{\partial n_p}{\partial r} + \frac{3}{2} n_p k_B \nu_{pe} (T_p - T_e) \nonumber\\
Q_e &=& \frac{3}{2} n_e u k_B \frac{\partial T_e}{\partial r} - u k_B T_e \frac{\partial n_e}{\partial r} - \frac{3}{2} n_e k_B \nu_{ep} (T_p - T_e) \nonumber \\ &+& \frac{1}{r^2} \frac{\partial}{\partial r} (q_{||,e} r^2 \text{cos}^2 \Phi).
\end{eqnarray}
The left hand sides are the volumetric heating rates of protons $Q_p$ and 
electrons $Q_e$, which can be determined from these 
equations if reasonable measures of the terms on the right hand sides are obtained. 
Note that among these terms we 
have proton-electron collisions of $\nu_{pe}$ and $\nu_{ep}$. We assume 
a constant outflow speed $u$ of 700 km s$^{-1}$. The Parker spiral angle is $\Phi$, given in its standard form as
\begin{equation} \label{eq:6}
\text{tan} \Phi = \Omega\,r\,\text{sin} \theta / u
\end{equation}
with a rotation frequency of $\Omega = $ 2.7 x 10$^{-6}$ rad s$^{-1}$ and a colatitude of $\theta = 15\degree$. We utilize the proton-electron collision rate scaling relations given by~\citet{Cranmer2009ApJ_empirical}:
\begin{equation} \label{eq:7}
\nu_{pe} \approx 8.4 \times 10^{-9} \left(\frac{n_e}{2.5\;\text{cm}^{-3}}\right)\left(\frac{T_e}{10^5\;\text{K}}\right)^{-3/2} \text{s}^{-1}
\end{equation}
\begin{equation} \label{eq:8}
\nu_{ep} \approx 8.4 \times 10^{-9} \left(\frac{n_p}{2.5\;\text{cm}^{-3}}\right)\left(\frac{T_p}{10^5\;\text{K}}\right)^{-3/2} \text{s}^{-1}
\end{equation}
The resulting values of heating rates are relatively insensitive to the choice of $u$ and $\theta$ in the inner heliosphere.

We numerically compute each derivative based on the fits in Equations~\ref{eq:1},~\ref{eq:2},~\ref{eq:3}, and~\ref{eq:4} using the standard centered-difference approximation on a grid of 1000 points.

\begin{figure}[t!]
    \centering
    \includegraphics[width=0.4\textwidth]{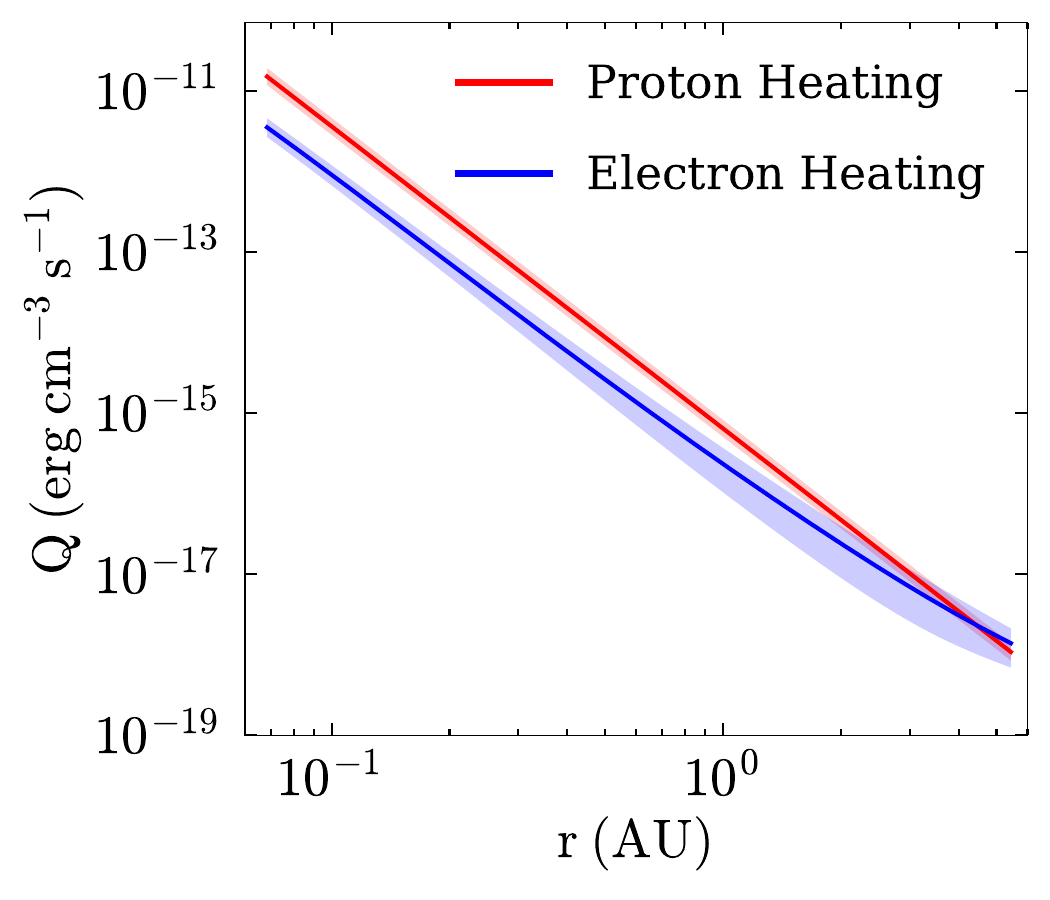} 
    \caption{Empirically derived heating rates vs. heliocentric distance. Also shown are the error envelopes described in the text.} \label{fig:2}
\end{figure}

\subsection{Results for Proton and Electron Heating} \label{subsec:results}
We solve Equations \ref{eq:5} for the proton and electron heating rates $Q_p$ and $Q_e$ over the range of heliocentric distances  covered by the first ten solar encounters by PSP, Helios, and Ulysses data. These datasets cover heliocentric distances ranging from 0.063 au to 5.44 au. Fig.~\ref{fig:2} shows the resulting heating rates. 

To estimate the errors corresponding to each computed 
heating rate, we construct the upper and lower limit of the proton and electron heating rates by the standard deviation of each quantity from  binning of the data. Then, for each quantity, the errors were propagated to the derived quantities. The relative uncertainties were calculated as nominal values across all radial distances. The obtained relative standard deviation values are $\delta_{T_p}=0.09$, $\delta_{T_e}=0.12$, $\delta_{q_{||,e}}=0.39$, and $\delta_{n_{e}}=0.15$. The uncertainty limits are shown by the translucent envelopes in Fig.~\ref{fig:2}. We find that the proton heating rate is well described by the following power-law fit, consistent with~\citet{Cranmer2009ApJ_empirical}, valid within 4\% relative accuracy across all distances:

\begin{equation} \label{eq:9}
Q_{p}\!=\!6.39\!\times\!10^{-16}\!\!\left(\frac{r}{\!1\;\text{au}}\right)^{\!\!-3.76}\!\!\left(\frac{u}{700\;\text{km s}^{-1}}\!\right) \!\text{erg}\;\text{s}^{-1}\text{cm}^{-3}
\end{equation}

This scaling relation is also comparable to the rate predicted by~\citet{VermaEA95} for Alfvénic streams of $Q_p \propto r^{-3.3}$. An approximate fit was obtained for the electron heating rate with around 51\% relative accuracy across all distances:

\begin{equation} \label{eq:10}
Q_{e}\!=\!2.70\!\times\!10^{-16}\!\!\left(\frac{r}{\!1\;\text{au}}\right)^{\!\!-3.27}\!\!\left(\frac{u}{700\;\text{km s}^{-1}}\!\right) \!\text{erg}\;\text{s}^{-1}\text{cm}^{-3}
\end{equation}

\begin{figure}
    \centering
    \includegraphics[width=0.4\textwidth]{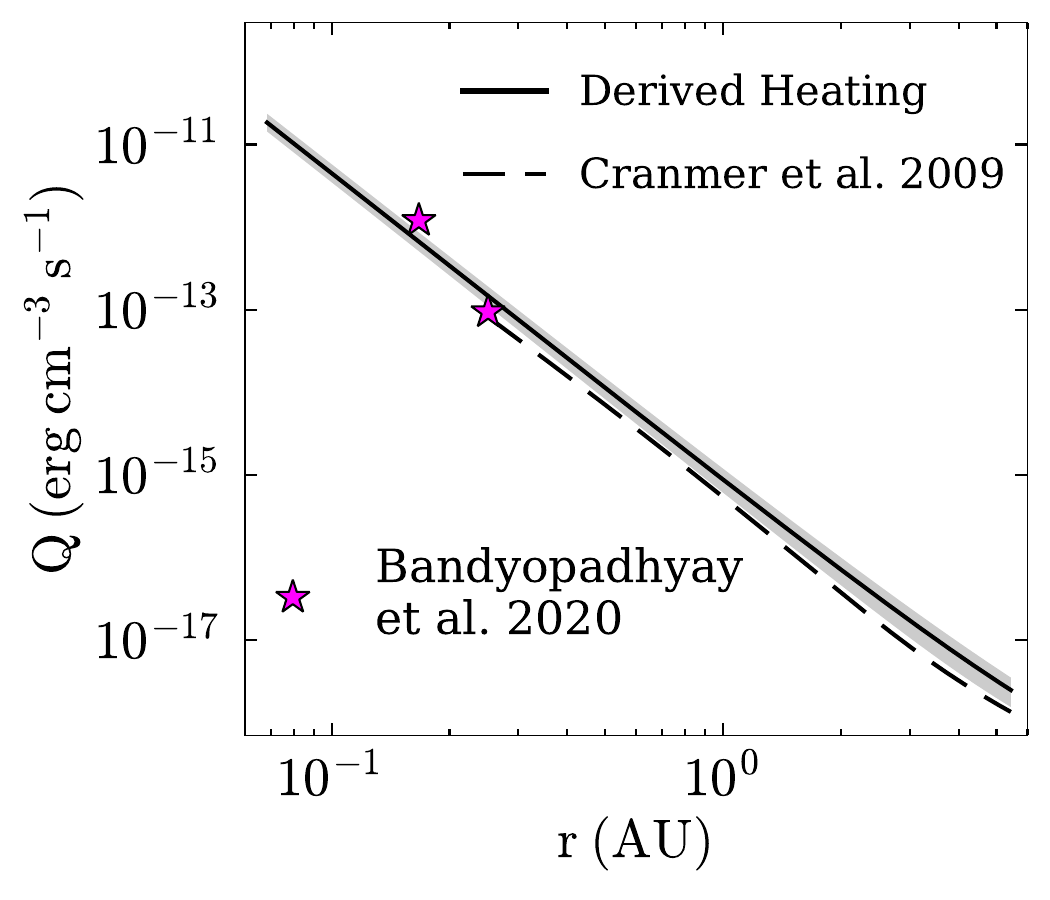}
    \caption{Total heating rates (proton plus electron) versus heliocentric distance. Error envelopes are described in text. The purple asterisks are the heating rate estimated using MHD cascade by \cite{Bandyopadhyay2020ApJS_cascade}.} \label{fig:3}
\end{figure}

We can also calculate the total heating rate and compare with previous results. The following fit is  valid to within 25\% relative accuracy:

\begin{equation} \label{eq:11}
Q_{tot}\!=\!9.25\!\times\!10^{-16}\!\!\left(\frac{r}{\!1\;\text{au}}\right)^{\!\!-3.58}\!\!\left(\frac{u}{700\;\text{km s}^{-1}}\!\right) \!\text{erg}\;\text{s}^{-1}\text{cm}^{-3}
\end{equation}

approximated by a power law fit $Q \propto r^{-\delta}$ with $\delta$ slightly greater than 3.58. The total heating is shown in Fig.~\ref{fig:3}. For comparison the estimate from \cite{Cranmer2009ApJ_empirical} is plotted. Also, the heating rate values estimated using the MHD Yaglom law, taken from \cite{Bandyopadhyay2020ApJS_cascade}, are plotted.

It is interesting to note that this behavior is consistent with the scaling expected from a von K\'arm\'an decay~\citep{Matthaeus2016ApJL}:
\begin{eqnarray} \label{eq:12}
    Q_p + Q_e \sim \rho \frac{Z^3}{\lambda},
\end{eqnarray}
where $\rho$ is the mass density, $Z$ is the turbulent amplitude and $\lambda$ is the correlation length.

This consistency can be readily demonstrated using even approximate canonical variations of the turbulence parameters. In the fast wind, usually $Z^2 \sim r^{-1}$ so $Z^3 \sim r^{-3/2}$. But $\lambda \sim  r^{1/2}$
and density approximately $\rho \sim r^{-2}$. Therefore, $\rho Z^3/\lambda$ is expected to scale as $r^{-4}$, which is close to the present estimates. More refined estimates of radial variations can be found in the literature~\citep[e.g.,][]{ChhiberEA21_large}. 
The same scaling was also observed by using a multi-spacecraft study following the same solar wind plasma at two different distances~\citep{Sorriso-Valvo2023AA_Helios}.

\begin{figure}
    \centering
    \includegraphics[width=0.4\textwidth]{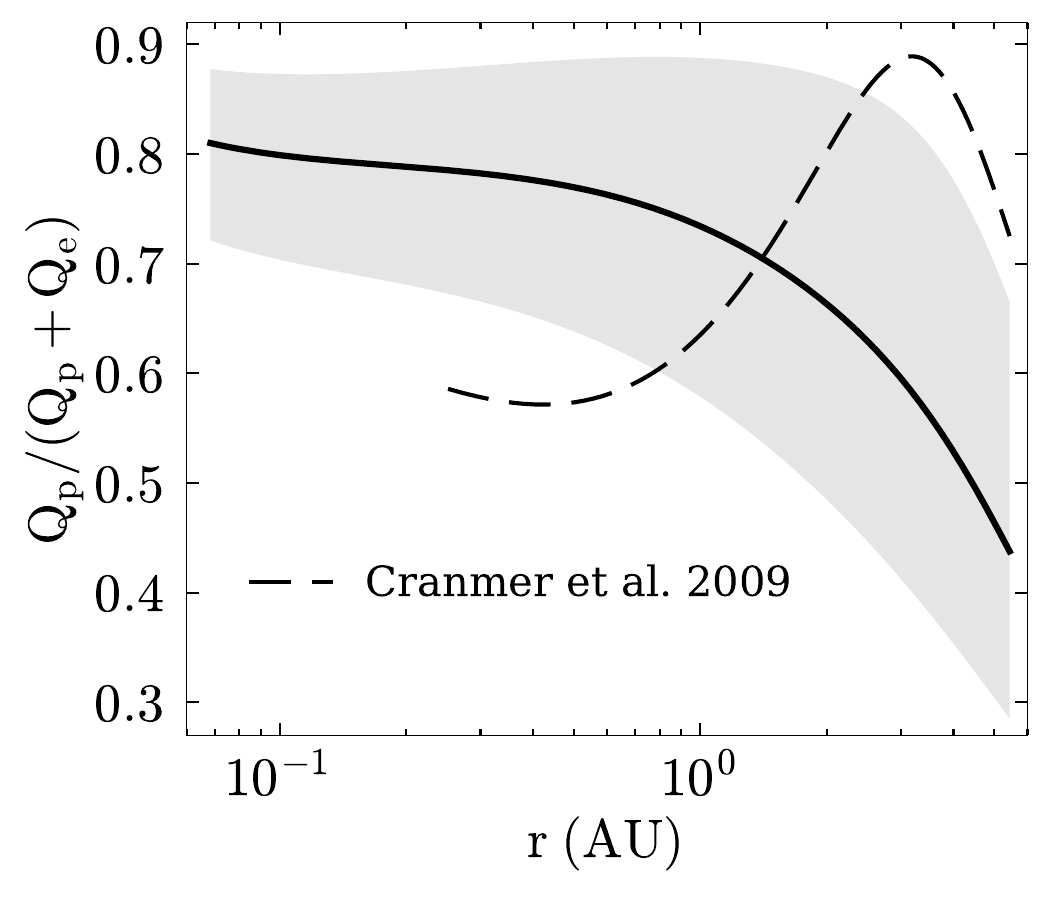}
    \caption{Proton heating to total heating ratio versus heliocentric distance. Uncertainty envelopes are described in text.} \label{fig:4}
\end{figure}

Perhaps the most interesting result of the paper is the estimation of heating fraction, plotted in Figure~\ref{fig:4}. The figure shows the ratio of the proton heating rate $Q_p$ to the total heating rate $Q_p + Q_e$, as a function of radial distance. The translucent gray envelope represents the uncertainty bar, evaluated by propagating the uncertainties from temperature, heat flux, and density. We notice a significant departure in behavior from that reported in~\citet{Cranmer2009ApJ_empirical}. The difference is mainly due to the difference in the fits to the electron heat conduction, as discussed before. 

\section{Conclusions and Discussion} \label{sec:conc}
This paper has computed empirical estimates for the rates of proton and electron heating in the fast solar wind from the near-Sun environment, from about 0.06 au to approximately 5 au. The PSP mission makes it possible to study these characteristics in the near-Sun solar wind, for the first time. The proton and electron heating rates were estimated from mean radial fits in the three data sets (as indicated by the least-squares fits shown in Fig.~\ref{fig:1} and given in equations~\ref{eq:1}-\ref{eq:4}), and they do not take into account the substantial spread that the data exhibits about the mean values. However, the results presented in this paper should give an estimation of the average heating fraction values for the radial distances covered.

A key aspect of this work has been the revised behavior of the electron heat conduction flux in the inner heliosphere, relative to the prior results by \cite{Cranmer2009ApJ_empirical}. This change is revealed only by inclusion of the PSP measurements. The slope of $q_{||, e}$ is slightly steeper than $r^{-2}$, leading to outward conduction of heat and local electron heating. Notably the new fits to the electron heat conduction flux significantly revise the heating partition trend at smaller radial distances~\citep[see][]{Cranmer2021JGR_heat-flux}. For example, we conclude that the protons receive about $70\%$ of the total plasma heating near 1 au, and this fraction increases to approximately $80\%$ at $< 15\,R_{\odot}$. At farther distances, near the orbit of Jupiter the proton heating fraction decreases to approximately $50\%$. Based on earlier analyses, modelers have frequently assumed that the protons receive about 60$\%$ of the dissipated energy over a wide range of distances. It is difficult to assess
at present what impact this revision might have on results of global modeling~\citep[such as][]{Usmanov2018ApJ,AdhikariEA17}, but it is possible that it will be significant. 

Solar wind heating has been shown to occur near inhomogeneously near coherent structures, which may accelerate energetic particles~\citep{Osman2012PRL, Bandyopadhyay2020ApJS_SEP}. Future works may explore how the heating rates of the different charged species change with the presence of energetic particles.

We recall that the heating rates $Q$ are calculated here 
by evaluating other relevant terms in the energy equations.
No form for $Q$ was prescribed and no specific mechanism was assumed. If the heating is due to turbulence, however, one would expect that $Q$ follows a von K\'arm\'an law. The present findings are indeed quite close to the scalings predicted using a von K\'arm\'an law, as explained previously. 
In this sense, the present results provide some justification for applicability of the use of the von K\'arm\'an MHD heating rate~\citep{Hossain1995PoF, Wan2012JFM, Bandyopadhyay2019JFM}
which is widely used in global solar wind MHD simulation codes~\citep[e.g.,][]{Usmanov2018ApJ}. However, we note that this is a very simplistic picture. Several plasma parameters, such as cross helicity, guide-field strength, and plasma beta might affect the scaling of Eq.~(\ref{eq:12}).

It is interesting to compare our estimates of the interplanetary proton and electron heating rates with those evaluated based on particular dissipation mechanisms. For example, a recent work by~\citet{Shankarappa2023arXiv_landau} estimates the proton and electron heating partition fraction using Landau damping. Although this work also suggests that protons are heated more than electrons, the proton heating fraction shows a slight decreasing trend at the inner heliosphere, close to the Sun. This is rather different from the behavior seen in Figure~\ref{fig:4}, which does not assume a particular dissipation mechanism. This contrast suggests that protons might be dominantly heated by other mechanisms although for electron dissipation, Landau damping might be the leading candidate~\citep{Chen2019NatCo_Landau, Afshari2021JGR_Landau}. 
Coronal observational data suggest that protons are preferentially heated over electrons
in the coronal holes~\citep{Kohl2006AAR_ultraviolet, Wilhelm2007SSR_ultraviolet, Landi2009ApJ_corona, Kasper2017ApJ_preferential}.
This observation appears to be compatible with 
the trend we have found that the proton heating 
fraction is greater at lower altitudes.

An improved understanding of the different charged species in the solar wind and coronal plasma is an important ingredient in characterizing heliospheric properties and predicting space weather events~\citep{Shi2023ApJ_temperatures}. Thus, inclusion of accurate proton and electron heating rates in global solar wind and space weather models may be a crucial step in improving their performance and predictive accuracy. This paper makes a key advance in that direction. 

\section{Acknowledgements}
Parker Solar Probe was designed, built, and is now operated by the Johns Hopkins Applied Physics Laboratory as part of NASA’s Living with a Star (LWS) program (contract NNN06AA01C). Support from the LWS management and technical team has played a critical role in the success of the Parker Solar Probe mission. We are deeply indebted to everyone that helped make the PSP mission possible. This research was partially supported by the Parker Solar Probe project through Princeton/IS$\odot$IS subcontract SUB0000165 and in part by PSP GI grant 80NSSC21K1767 at Princeton University. Support is also 
acknowledged to the Heliospheric Supporting Research program grants
80NSSC18K1210 and 80NSSC18K1648, and a Parker Solar Probe Guest Investigator program 80NSSC21K1765 at Delaware.

This research was supported by the International Space Science Institute (ISSI) in Bern, through ISSI International Team project $\#$560.

The authors acknowledge CNES (Centre National d\'Etudes Spatiales), CNRS (Centre National de la Recherche Scientifique), the Observatoire de PARIS, NASA, and the FIELDS/RFS team for their support to the PSP/SQTN data production, and the CDPP (Centre de Donnees de la Physique des Plasmas) for their archiving and provision.

Ion density data from \textit{Helios} and \textit{Ulysses} were extracted directly from the National Aeronautics and Space Administration's online archive \url{https://spdf.gsfc.nasa.gov/data\_orbits.html}. 

\appendix

\section{Comparison with Previous Fits}\label{sec:pic}
\begin{figure}[ht!]
    \centering
    \includegraphics[width=0.35\columnwidth]{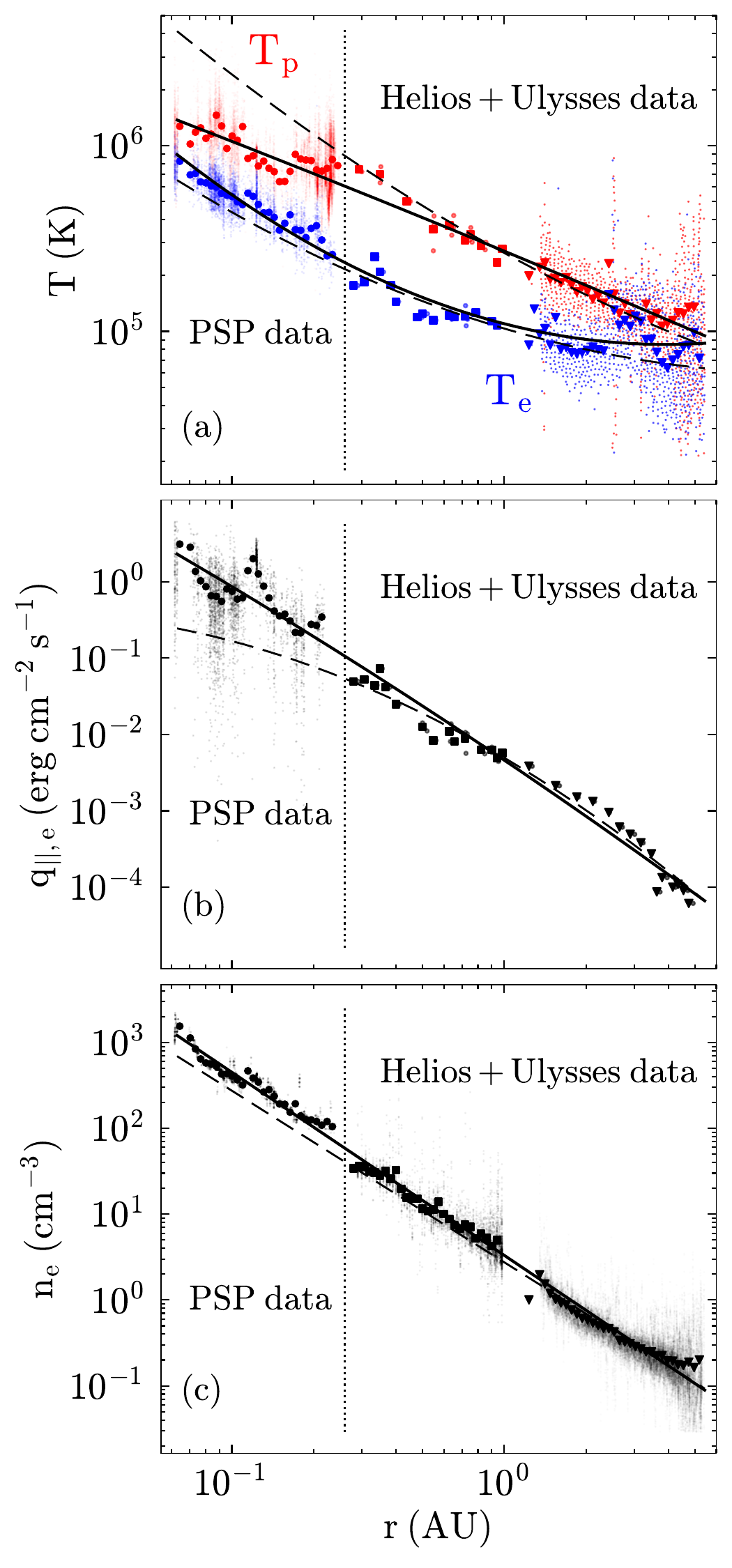}
    \caption{Our fits are shown in solid black lines, and the previous fits from \cite{Cranmer2009ApJ_empirical} are plotted in dashed black lines.}
    \label{fig:comp}
\end{figure}

Here we show the comparison between the fits of the various data obtained in this paper with those from the previous study by \cite{Cranmer2009ApJ_empirical}. Fig.~\ref{fig:comp} shows the earlier fits in dashed lines. Although there is some deviation between the fits in each variable, the most prominent difference between the old and new fits can be seen for parallel heat conduction flux.

\clearpage

\bibliographystyle{aasjournal}
 \newcommand{\BIBand} {and} 
\newcommand{\boldVol}[1] {\textbf{#1}} 
\providecommand{\SortNoop}[1]{} 
\providecommand{\sortnoop}[1]{} 
\newcommand{\stereo} {\emph{{S}{T}{E}{R}{E}{O}}} 
\newcommand{\au} {{A}{U}\ } 
\newcommand{\AU} {{A}{U}\ } 
\newcommand{\MHD} {{M}{H}{D}\ } 
\newcommand{\mhd} {{M}{H}{D}\ } 
\newcommand{\RMHD} {{R}{M}{H}{D}\ } 
\newcommand{\rmhd} {{R}{M}{H}{D}\ } 
\newcommand{\wkb} {{W}{K}{B}\ } 
\newcommand{\alfven} {{A}lfv{\'e}n\ } 
\newcommand{\alfvenic} {{A}lfv{\'e}nic\ } 
\newcommand{\Alfven} {{A}lfv{\'e}n\ } 
\newcommand{\Alfvenic} {{A}lfv{\'e}nic\ }

\end{document}